\documentclass[%
 reprint,
superscriptaddress,
floatfix,
 amsmath,amssymb,
 aps,
]{revtex4-1}

\usepackage[utf8]{inputenc}
\usepackage{amsmath,esint}
\usepackage{mathtools}
\usepackage[colorlinks=true,linkcolor=blue,citecolor=blue,urlcolor=blue]{hyperref}
\usepackage{amsfonts}
\usepackage{amssymb}
\usepackage{bm}
\usepackage{textcomp}
\usepackage{empheq}
\usepackage{siunitx}
\usepackage[titletoc,title]{appendix}
\usepackage{graphicx}% Include figure files
\usepackage{dcolumn}% Align table columns on decimal point
\usepackage{bm}% bold math
\usepackage{subcaption}
\usepackage{xcolor,soul}

\renewcommand{\Im}{\operatorname{Im}}

\DeclareMathOperator{\Tr}{Tr}
\newcommand{\citeasnoun}[1]{Ref.~\cite{#1}}

\newcommand{\Figref}[1]{Figure~\ref{fig:#1}}
\newcommand{\figref}[1]{Fig.~\ref{fig:#1}}
\renewcommand{\eqref}[1]{Eq.~(\ref{eq:#1})}
\newcommand{\Eqref}[1]{Equation~(\ref{eq:#1})}

\newcommand{\vect}[1]{\boldsymbol{\mathbf{#1}}}

\newcommand{\secref}[1]{Sec.~\ref{sec:#1}}

\newcommand*{\Ev}{\mathbf{E}}

\newcommand*{\xv}{\mathbf{x}}
\newcommand*{\Pv}{\mathbf{P}}

\newcommand*{\Escat}{\Ev_{\rm scat}}

\newcommand*{\Einc}{\Ev_{\rm inc}}

\begin{document}

\preprint{APS/123-QED}

\title{Fundamental limits to multi-functional and tunable nanophotonic response}

\author{Hyungki Shim}
\thanks{These authors contributed equally to this paper.}
\affiliation{Department of Applied Physics and Energy Sciences Institute, Yale University, New Haven, Connecticut 06511, USA}
\affiliation{Department of Physics, Yale University, New Haven, Connecticut 06511, USA}
\author{Zeyu Kuang}
\thanks{These authors contributed equally to this paper.}
\affiliation{Department of Applied Physics and Energy Sciences Institute, Yale University, New Haven, Connecticut 06511, USA}
\author{Zin Lin}
\affiliation{Department of Mathematics, Massachusetts Institute of Technology, Cambridge, Massachusetts 02138, USA}
\author{Owen D. Miller}
\email{owen.miller@yale.edu}
\affiliation{Department of Applied Physics and Energy Sciences Institute, Yale University, New Haven, Connecticut 06511, USA}

\date{\today}

\begin{abstract}
    Tunable and multi-functional nanophotonic devices are used for applications from beam steering to sensing. Yet little is understood about fundamental limits to their functionality. The difficulty lies with the fact that it is a single structure that must exhibit optimal response over multiple scenarios. In this article, we present a general theoretical framework for understanding and computing fundamental limits to multi-functional nanophotonic response. Building from rapid recent advances in bounds to light-matter interactions, we show that after rewriting the design problems in terms of polarization fields, the introduction of suitable cross-correlation constraints imposes the crucial ``single-structure'' criteria. We demonstrate the utility of this approach for two applications: reflectivity contrast for optical sensing, and maximum efficiency for optical beam switching. Our approach generalizes to any active or multi-functional design in linear optics.
\end{abstract}

\pacs{Valid PACS appear here}% PACS, the Physics and Astronomy
\maketitle

\section{Introduction}
Nanophotonic devices that offer multiple functionalities, from liquid-crystal devices for beam steering~\cite{Resler1996,Beeckman2011,Li2019,He2019} to polychromatic metasurface lenses~\cite{Khorasaninejad2015,Avayu2017,Chen2017,Chen2018,Paniagua-Dominguez2018,Wang2018,Shrestha2018,Chung2020}, have tremendous design complexity due to the need for a single structure to simultaneously optimize multiple objectives. In this Letter, we develop a general framework for identifying fundamental limits to achievable response in multi-functional devices. Despite significant recent interest in identifying nanophotonic bounds~\cite{Miller2016,Miller2015,Hugonin2015,Miller2017a,Yang2017,Yang2018,Shim2019,Michon2019,Zhang2019,Molesky2019a,Shim2020a,Shim2020,Gustafsson2020,Kuang2020b,Molesky2020b,Shim2021,Miller2021}, such works have almost exclusively applied only to devices with only a single function. Here, building from recent discoveries that design problems can be transformed to quadratically constrained quadratic programs~\cite{Kuang2020b,Molesky2020c}, we introduce a simple mechanism for constructing ``cross-correlation'' constraints that encapsulate simultaneous requirements on a single device. Such constraints can be utilized for maximum functionality across multiple frequencies, incident fields, and constituent-material properties, as well as active modulation. We identify two prototypical examples where such bounds illuminate the limits to what is possible: multi-frequency reflection control in optical filters, for applications such as optical sensing, and optimal beam switching via liquid-crystal-based nanophotonic metagratings. Our framework promises to identify the limits to complex optical functionality for wide-ranging applications.

Fundamental limits to optical response have dictated technological selection for decades, ranging from the Wheeler--Chu limits to broadband antenna design for mobile devices~\cite{Wheeler1947,Chu1948,Sievenpiper2012,Pfeiffer2017} to photovoltaic energy-conversion efficiency approaching the canonical Shockley--Queisser limits~\cite{Shockley1961,Guillemoles2019,Swanson2005,Miller2012}. Similarly, imaging techniques are measured against the Abbe diffraction limit, while all-angle, broadband absorbers are designed against the Yablonovitch limit~\cite{Yablonovitch1982a}. Yet none of these approaches account for the full complexity of wave-scattering physics allowed by Maxwell's equations, instead utilizing small-size (Wheeler--Chu), free-space propagation (Abbe), or ray-optical (Yablonovitch) regimes in which the physics is dramatically simpler. 

Recent progress has suggested the possibility for identifying bounds in more complex, ``full-wave'' regimes. After developing the idea of ``communication channels,'' which limit information capacities transmitted between known scattering bodies~\cite{Miller2000,miller2012all,Miller2019}, D. A. B. Miller utilized single-frequency scattering sum rules to identify bounds on corralling and separating pulses, and more complex functionality~\cite{Miller2007,Miller2008a}. Those bounds, however, take specialized consideration of currents non-orthogonal to ``pass-through'' and ``single-pass'' waves in a way that appears difficult to extend beyond one-dimensional structures. All-frequency sum rules utilize complex-frequency contour integrals to connect all-frequency response to relatively simple constants (related to either electrostatic response or electron densities)~\cite{gordon_1963, purcell_1969, mckellar_box_bohren_1982,sohl_gustafsson_kristensson_2007,Gustafsson2010,Bernland2010,Yang2015,Cassier2017,Shim2019}, but can only be utilized for very special response functions such as extinction, and they cannot meaningfully bound response over smaller frequency ranges. 

Analytical bounds to single-frequency response have been developed using a variety of constraints that essentially distill to energy-conservation constraints over the relevant scattering bodies~\cite{Yaghjian1996,hamam_coupled-mode_2007,kwon2009optimal, liberal2014least, liberal2014upper, Miller2016,Miller2015,Hugonin2015,Miller2017a,Yang2017,Yang2018,Michon2019,Molesky2019a,Gustafsson2020,Miller2021}. One can accommodate multiple functionalities in the sense of, for example, maximizing scattering while constraining absorption~\cite{Yang2017,Miller2021}, but multiple scenarios cannot be accounted for. It is possible to combine the contour-integral, broad-bandwidth approach with the energy-conservation-based, single-frequency approach to identify bounds over any frequency range~\cite{Shim2019}, but such bounds are feasible only for response functions with very special optical-theorem-like structure, such as local density of states~\cite{Shim2019} and extinction~\cite{Kuang2020b}. Tighter single-frequency bounds can be achieved with a recently developed computational approach in which many ``local'' conservation laws are identified that must necessarily be satisfied by any design~\cite{Kuang2020b,Molesky2020c}. The mathematical framework of this approach appears to generalize to physical design problems across many domains; one successful application has been to quantum optimal control~\cite{Zhang2021}. A unique instance of multiple functionalities, that of multiple incident fields (or initial wavefunctions in the quantum case) was proposed in Refs.~\cite{Molesky2021,Zhang2021}, while an alternative formulation of multi-scenario design was proposed in \citeasnoun{Angeris2021}. However, the approach of \citeasnoun{Angeris2021} to multiple scenarios does not actually incorporate any constraints between scenarios, and hence the bounds for many scenarios simplify to a weighted average of the single-scenario bounds.

In this article, we combine the recently developed integral-equation framework for optical-response bounds~\cite{Kuang2020b,Molesky2020c} with a new approach to capturing constraints that must be satisfied in multi-functional devices. The key is to identify generalized conservation laws that relate the polarization fields induced in one scenario to the incident and scattered fields in another (\secref{genform}). These conservation laws are necessary \emph{and} sufficient conditions for any design solution, and therefore one can replace the Maxwell-equation constraints, which are nonlinear (and non-polynomial) in the structural design variables, with a set of quadratic constraints in the polarization fields. From this quadratically constrained quadratic program (QCQP), one can ``lift'' the degrees of freedom to a higher-dimensional space, and relax the constraints, such that the problem becomes a semidefinite program (SDP) whose global optimum can be computed efficiently~\cite{Nocedal2006}. \Eqref{QCQP} represents the culmination of the transformations to a QCQP; the solutions of the corresponding SDP represent fundamental limits to what is possible is multi-functional systems. We apply this approach to two classes of applications in \secref{multiappl}, where we show that the limits illuminate fundamental constraints on traits such as size, sensitivity, optical contrast, and switching efficiency in multi-functional devices.

\section{A Framework for Multi-Functional Bounds} \label{sec:genform}
We start with a conventional statement of the multi-functional Maxwell design problem, of maximizing some objective subject to the Maxwell-equation constraints. We show that the problem can be transferred to one with a different mathematical structure, a QCQP, without any approximation or relaxation, through identification of power-conservation and cross-correlation constraints embedded in the structure of Maxwell's equations. Given the QCQP formulation, one can ``relax'' the problem through now-standard techniques (e.g. semidefinite relaxations), which leads to problems that are efficiently solvable and whose solutions represent global bounds on what is possible.

In a Maxwell design problem, one maximizes (or minimizes) some objective subject to the constraint of Maxwell's equations over all scenarios of interest. For notational simplicity, we will assume here nonmagnetic, isotropic materials in free space. (The generalization to anisotropic materials in arbitrary backgrounds is straightforward). For $N$ scenarios of interest (incident fields, frequencies, etc.), if we denote the field of scenario $s$ as $\Ev_s(\xv)$, and take the designable geometry as a spatially dependent electric susceptibility $\chi(\xv)$, then one can mathematically write the design problem as
\begin{equation}
    \begin{aligned}
        & \underset{\chi_s(\xv)}{\text{max.}} & & f(\Ev_s(\xv)) \\
        & \text{s.t.} & & 
        \left[\nabla \times \nabla \times - (1 + \chi_s(\xv))\omega_s^2\right] \Ev_s(\xv) = i\omega \vect{J}_s(\xv),
    \end{aligned}
    \label{eq:MaxDesProb}
\end{equation}
for all $s$ from $1$ through $N$, where we have taken $e^{-i\omega_s t}$ time-dependencies and units $\varepsilon_0 = \mu_0 = 1$. The key mathematical difficulty in solving such design problems is that the Maxwell equations are nonconvex in the geometrical degrees of freedom $\chi_s(\xv)$; in fact, the design problem is in the class of NP hard problems~\cite{Angeris2021}. We will show, however, that under suitable transformations, bounds can be identified via efficient algorithms.

Our first step is to transform the differential form of \eqref{MaxDesProb} to an equivalent integral-equation form, which is then amenable to the transformation to a QCQP. The volume-integral formulation arises by separating the total fields $\Ev$ into its incident and scattered components, $\Einc$ and $\Escat$, respectively, and then re-writing the total and scattered fields instead in terms of the polarization currents that generate them. First, the definition of susceptibility implies that $\Ev(\xv) = \chi^{-1}(\xv) \Pv(\xv)$, for polarization field $\Pv$ and at any point $\xv$ in the material (such that $\chi(\xv) \neq 0$). Second, the scattered field is the field radiated by the polarization currents as though they were in free space, which is given by the convolution of the free-space (or background) dyadic Green's function with the polarization field: $\Escat(\xv) = \int_V G_0 (\xv,\xv') \Pv(\xv')$, where $V$ is the volume of the designable region. Hence one can replace the differential form of Maxwell's equation in \eqref{MaxDesProb} with the equivalent integral equations~\cite{Chew2008},
\begin{align}
    \left[\chi_s^{-1}(\xv) \Pv_s(\xv) - \int_V G_{0,s}(\xv,\xv') \Pv_s(\xv') \,{\rm d}\xv'\right] = \Ev_{\textrm{inc},s}(\xv),
    \label{eq:IntDesProb}
\end{align}
again over all scenarios $s$. Again, however, the expressions are nonconvex and hard to work with due to the presence of both variables $\chi(\xv)$ and $\Pv(\xv)$ in the expressions. Also, \eqref{IntDesProb} is only valid at points within the material for which $\chi(\xv) \neq 0$, which is an unknown domain before the design problem has been solved.

We can simplify the integral constraint into a simpler quadratic form that extends to all design space $V$ through the following idea. In the designable region, the desired material is either present or absent. If the material is absent, \eqref{IntDesProb} simplifies to $\Pv_s(\xv) = 0$, as there is no polarization field. If the material is present, then at that point we know $\chi_s(\xv)$ equals the \emph{constant} value $\chi_s$, without position argument, and \eqref{IntDesProb} simplifies to a linear equation in $\Pv_s$. Hence at any point $\xv$ in the design region, either $\Pv_s(\xv) = 0$ (in the absence of material), or $\chi_s^{-1} \Pv_s(\xv) - \int_V G_{0,s}(\xv,\xv') \Pv_s(\xv') \,{\rm d}\xv' - \Ev_{\textrm{inc},s}(\xv) = 0$ (in the presence of material). We can combine these possibilities into a single, quadratic constraint, by multiplying them together (and conjugating the first equation for mathematical convenience):
\begin{align}
    \Pv^*_s(\xv) &\cdot \left[\chi_s^{-1} \Pv_s(\xv) - \int_V G_{0,s}(\xv,\xv') \Pv_s(\xv') \,{\rm d}\xv' - \Ev_{\textrm{inc},s}(\xv)\right] \nonumber \\
    & = 0.
    \label{eq:QuadConstr}
\end{align}
If we enforce the constraint of \eqref{QuadConstr} at all points $\xv$ in the designable region, it is equivalent to enforcing the integral (or differential) Maxwell constraints. One can show that the real and imaginary parts of \eqref{QuadConstr} represent conservation of real and reactive power flow through an infinitesimal bounding surface centered at $\xv$. These constraints (or equivalent versions of them~\cite{Molesky2019a}) were used in Ref.~\cite{Kuang2020b} to identify bounds in single-function photonic structures. Their key drawback, however, is that they do not enforce any restrictions between different scenarios. For example, if one wants a single structure to exhibit optimal response at two frequencies, there is no restriction in \eqref{QuadConstr} that the polarization fields in the two scenarios must arise from the same structure. Hence, such bounds (including those discussed in ~\citeasnoun{Angeris2021}) yield multi-functional design bounds that are often trivial in some way.

\subsection{Cross-correlation constraints}
\begin{figure} [t!]
    \includegraphics[width=1\linewidth]{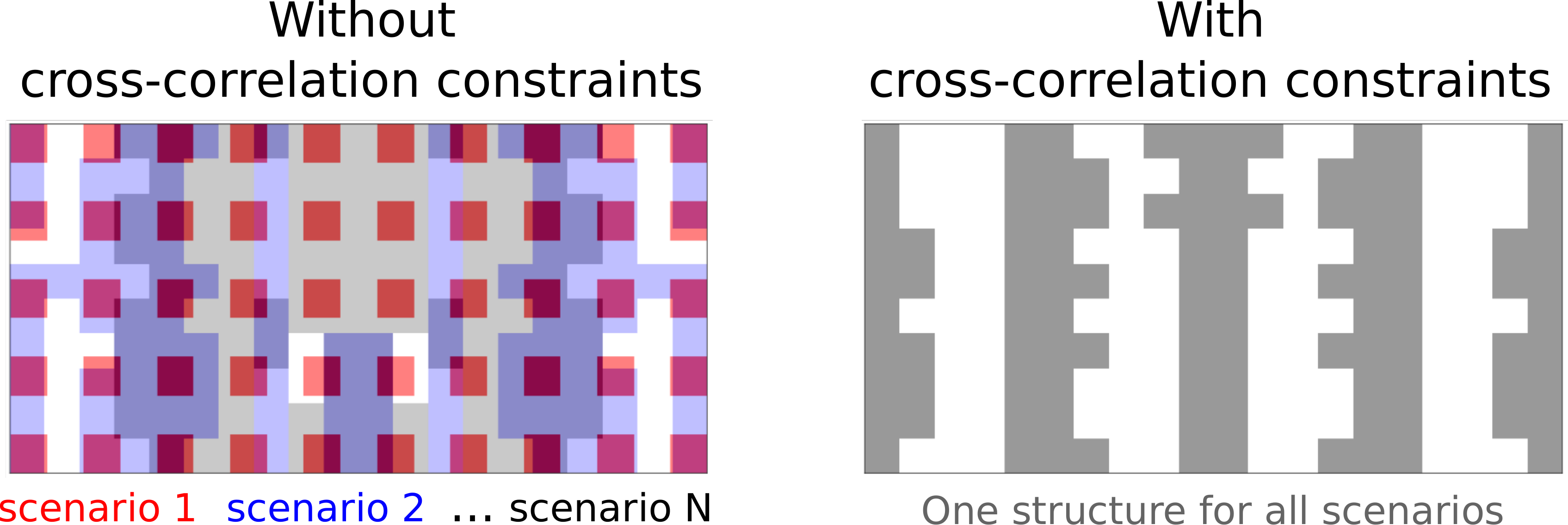} 
    \caption{Previous approaches to identifying fundamental limits could not account for the constraint that a single device must account for response across all scenarios with a multi-functional objective (left). We introduce cross-correlation constraints in \eqref{CCConstr} that correctly account for a single device (right), yielding meaningful bounds on what is possible.}  
    \label{fig:crosscorr} 
\end{figure}

Here, we show that one can naturally enforce the desired constraints through the creation of new, ``cross-correlation'' constraints. The key idea is to consider the two terms multiplied together in \eqref{QuadConstr}, and to take similar products of the terms \emph{across all combinations of different scenarios}. Mathematically, this means adding constraints of the form
\begin{align}
\Pv^*_{s_i}(\xv) \cdot \bigg[\chi_{s_j}^{-1} \Pv_{s_j}(\xv) &- \int_V G_{0,s_j}(\xv,\xv') \Pv_{s_j}(\xv') \,{\rm d}\xv'  \nonumber \\
                                                                & - \Ev_{\textrm{inc},s_j}(\xv)\bigg] = 0,
    \label{eq:CCConstr}
\end{align}
for all combinations of scenarios, $i \neq j$. Why does this enforce the desired constraints? Consider a given $i$ and $j$. \Eqref{CCConstr} requires that either the material in scenario $i$ is absent (hence $\Pv_{s_i} = 0$), or the material in scenario $j$ is present. This achieves half of the necessary requirement: if the material in scenario $i$ is present, and hence the first condition fails, then the material in scenario $j$ must also be present. The reverse condition, that the absence of one material must imply the absence of the other, arises from the equivalent set of constraints in which $i$ and $j$ are reversed. \Figref{crosscorr} illustrates the difference between imposing the cross-correlation constraints and not: without the cross-correlation constraints, different optimal structures would implicitly be considered for each scenario. Using the cross-correlation constraints, one can enforce a single structure that must be utilized for each scenario. 
	 
\subsection{Optimization problem} %(write down the general form of objective and constraints)
The transformations discussed above lead us to a new formulation of the Maxwell design problem, in terms of the polarization fields for all scenarios, $\Pv_s(\xv)$:
\begin{equation}
    \begin{aligned}
        & \underset{\Pv_s(\xv)}{\text{max.}} & & f(\Pv_s(\xv)) \\
        & \text{s.t.} & & \textrm{\Eqref{CCConstr} satisfied for all $i,j$} \\
        & & & \textrm{ and all $\xv$ in the design region.}
    \end{aligned}
    \label{eq:QCQP}
\end{equation}
Any typical objective $f$ of interest in linear optics (scattered/absorbed power, spontaneous-emission enhancement, mode overlap, etc.) will be a quadratic (or linear) function of the polarization fields $\Pv_s$, in which case \eqref{QCQP} is a quadratic objective subject to quadratic constraints, and is thereby a QCQP. Formulating a problem as a QCQP does not necessarily make it easy to solve; generically, QCQPs are NP hard to solve~\cite{Luo2010}. However, they arise in wide-ranging problems across engineering and physics~\cite{Williamson1994,Vandenberghe1996,Hongwei2004,Biswas2006,Gershman2010,Luo2010}, and numerous tools have been developed to aid in their solution. In particular, \emph{semidefinite relaxations} provide a mechanism for relaxing QCQPs to semidefinite programs (SDPs), which can be solved efficiently using interior-point methods~\cite{Boyd2004}.

The general prescription to relax a QCQP is to ``lift'' the variable to a higher-dimensional space in which the quadratic constraints are transformed to linear constraints~\cite{Luo2010}. In our case, let us assume that we stack the space-, polarization-, and scenario-dependent polarization-field values into a single vector $\vect{p}$. Then our constraints are of the form $\vect{p}^\dagger \vect{A} \vect{p} + \vect{y}^\dagger \vect{p} = 0$. The difficult part is the nonconvex quadratic term $\vect{p}^\dagger \vect{A} \vect{p}$. That term can be re-written as $\Tr\left(\vect{A} \vect{p} \vect{p}^\dagger\right)$; if we define a new matrix variable $\vect{X} = \vect{p}\vect{p}^\dagger$, then the constraint now has the linear form $\Tr\left(\vect{A}\vect{X}\right) + \vect{y}^\dagger \vect{x}$. The variable $\vect{X}$ is not free to take any form, however, as it must be a rank-one positive semidefinite matrix. The rank-one constraint is the only nonconvex constraint, and it can be dropped to transform the problem to a convex SDP. This is the only necessary relaxation, and it is the point at which any looseness (e.g. non-enforcement of the single-structure constraint) may arise. Finally, the corresponding SDP can be solved using standard packages~\cite{cvx,gb08}, and the resulting solution is a fundamental limit on the multi-functional design problem of interest.

\section{Applications} \label{sec:multiappl}
     
To illustrate the generality and utility of the bound framework established in \secref{genform}, we demonstrate its ability to predict fundamental limits for two prototype problems: the design of maximum-reflectivity-contrast optical filters, as may be useful for sensing applications, and the design of a liquid-crystal-based beam-switching device. In each case we show that previous bounds trivialize in some way, whereas our bounds can be quite close to (and always above) the performance of real theoretical designs. These applications demonstrate that our bounds identify fundamental limits to what is possible, and that they tease out the key scaling laws of multi-functional systems.

\subsection{Maximum reflectivity contrast} \label{sec:reflcontrast}

\begin{figure} [t!]
    \includegraphics[width=0.9\linewidth]{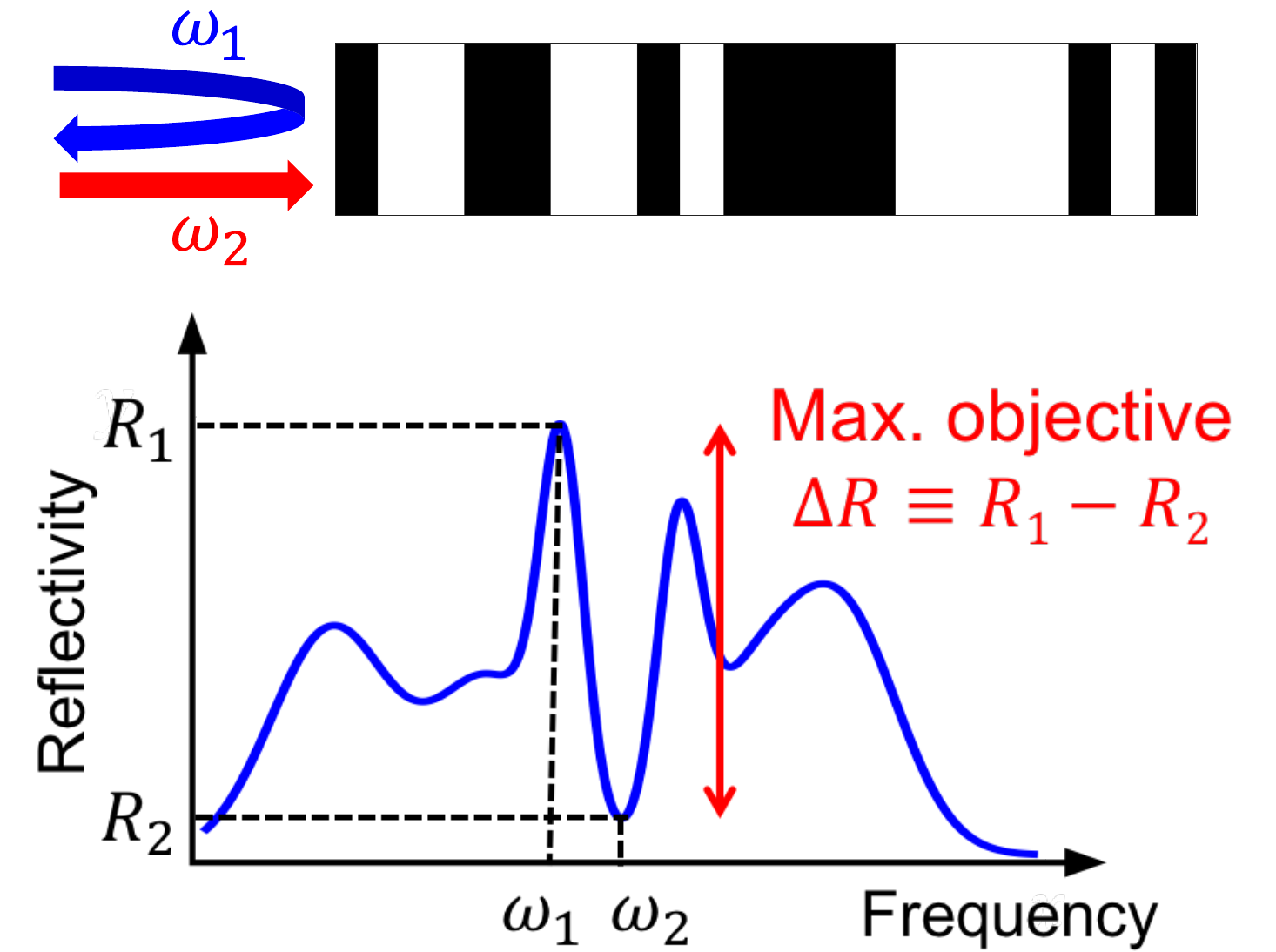} 
    \caption{The first application we consider is that of maximum reflectivity contrast between two nearby frequencies, $\omega_1$ and $\omega_2$, on a multilayer structure. Our bounds capture tradeoffs between the key parameters such as designable-domain size, refractive index, frequency separation, and reflectivity contrast.}  
    \label{fig:reflscheme} 
\end{figure} 	
			
A key problem in sensing applications~\cite{Anker2009b,Chang2010,Liu2010,Law2013a} is to detect changes, however minuscule, in the incident frequency of light. To do so, the scatterer must be designed to maximize the variation in its response with respect to changes in incident frequency. While there are several candidates for such optical response, we focus here on reflectance, which naturally arises in many applications~\cite{Liu2010,Law2013a}, and on multilayer films, i.e., optical filters. The problem is depicted schematically in \figref{reflscheme}, with any combination of angles and frequencies for the incident, reflected, and transmitted waves, while the primary objective might be the normal-incident contrast in reflectivity between two nearby frequencies. We take the multilayer film to comprise alternating layers of Al$_2$O$_3$ (as is commonly used for multilayer thin films~\cite{Dobrowolski2008,Poitras2017}) and vacuum. Reflectance is a quadratic function of the polarization fields. Specifically, the reflection coefficient is the amplitude of the back-scattered wave, which can be evaluated through a convolution of the free-space Green's function with the polarization field, and the reflectance $R(\omega)$ is the square of this quantity. Our objective is the contrast in reflectivity at any two frequencies of interest, which we can write as $R(\omega_1) - R(\omega_2)$. Our set of constraints is those of \eqref{QCQP}, including the cross-correlation constraints between the polarization fields at the two frequencies. 

\begin{figure*} [t!]
    \includegraphics[width=1\linewidth]{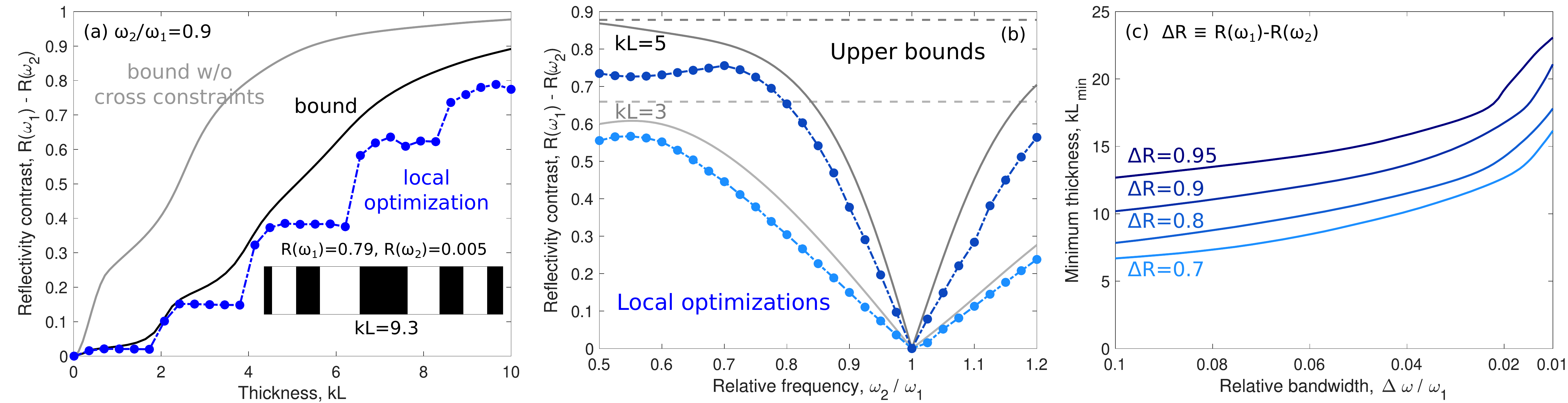} 
    \caption{(a) Comparison of bounds on reflectivity contrast, both with and without cross-correlation constraints, to designs achieved via gradient-based local optimization. The relative frequency $\omega_2 / \omega_1$ is set to 0.9, and $kL$ is the normalized thickness of the designable region $k=2\pi / \lambda$. The inset shows a locally-optimized design at $kL=9.3$, with reflectance of 0.79 at $\omega_1$ and 0.005 at $\omega_2$. (b) Similar to (a), but as a function of relative frequency $\omega_2 / \omega_1$, where $kL$ is fixed at 3 and 5. (c) The minimum thickness required for different values of desired reflectivity contrast, as predicted by our bounds with cross-correlation constraints. Since reflectance ranges from 0 to 1, the reflectivity contrast ranges from -1 to 1.}  
    \label{fig:RvsL} 
\end{figure*} 
			
The solid black lines in \figref{RvsL}(a) show the global bounds for reflectivity contrast as a function of the maximum thickness $L$ of the designable region (normalized to wave number $k = \omega / c$). Also included are the bounds without the cross-correlation constraints, as well as actual reflectivity contrasts achieved by designs identified through a local-optimization routine. For each thickness of the designable region, we used gradient descent, a standard optimization technique~\cite{Nocedal2006}, to optimize over many initial conditions (random designs) and select the best-performing design. In each case, one can see that the results of the local optimizations are able to quite closely approach, though not surpass, the global bounds. This suggests that the bounds have very little ``slack,'' i.e., are nearly the best possible bounds. The bounds without the cross-correlation constraints, shown in grey, are not as close to the local-optimization results. In fact, the bounds without the cross-correlation constraints are trivial in the sense that they always identify zero reflectance as possible for $R(\omega_2)$ (since an all-vacuum structure would be reflectionless), and there are no constraints requiring the same structure to also provide large reflectance at a nearby frequency. One can see that for two relatively close frequencies, it is vital to incorporate the cross-correlation constraints to achieve meaningful bounds.

The inability of previous bound approaches to identify key features of multi-functional bounds is further illustrated in \figref{RvsL}(b), which sweeps across a variety of relative frequency values (holding $\omega_1$ fixed). In each case, the solid line depicts the global bound, which tracks quite closely with the best designs from the local-optimization computations. By contrast, the dashed lines show the bounds without the cross-correlation constraints, which trivially do not change as a function of relative frequency because $R(\omega_2)$ is always bounded below by 0. Using the cross-correlation constraints is crucial to capturing the difficulty of achieving high contrast for small changes in frequency. 

Finally, we can use this approach to identify a bound on the minimum thickness of any design that could possibly achieve a desired reflectivity contrast $\Delta R$ as a function of the relative bandwidth of the two frequencies of interest, as shown in \figref{RvsL}(c). As expected, the minimum thickness increases both as a function of the desired reflectivity difference as well as as a function of decreasing relative bandwidth. But the scaling lines of \figref{RvsL}(c) that indicate precise tradeoffs between sensitivity and size would not be possible with any other approach.

\subsection{Beam switching via liquid crystals}

\begin{figure*} [t!]
    \includegraphics[width=0.8\linewidth]{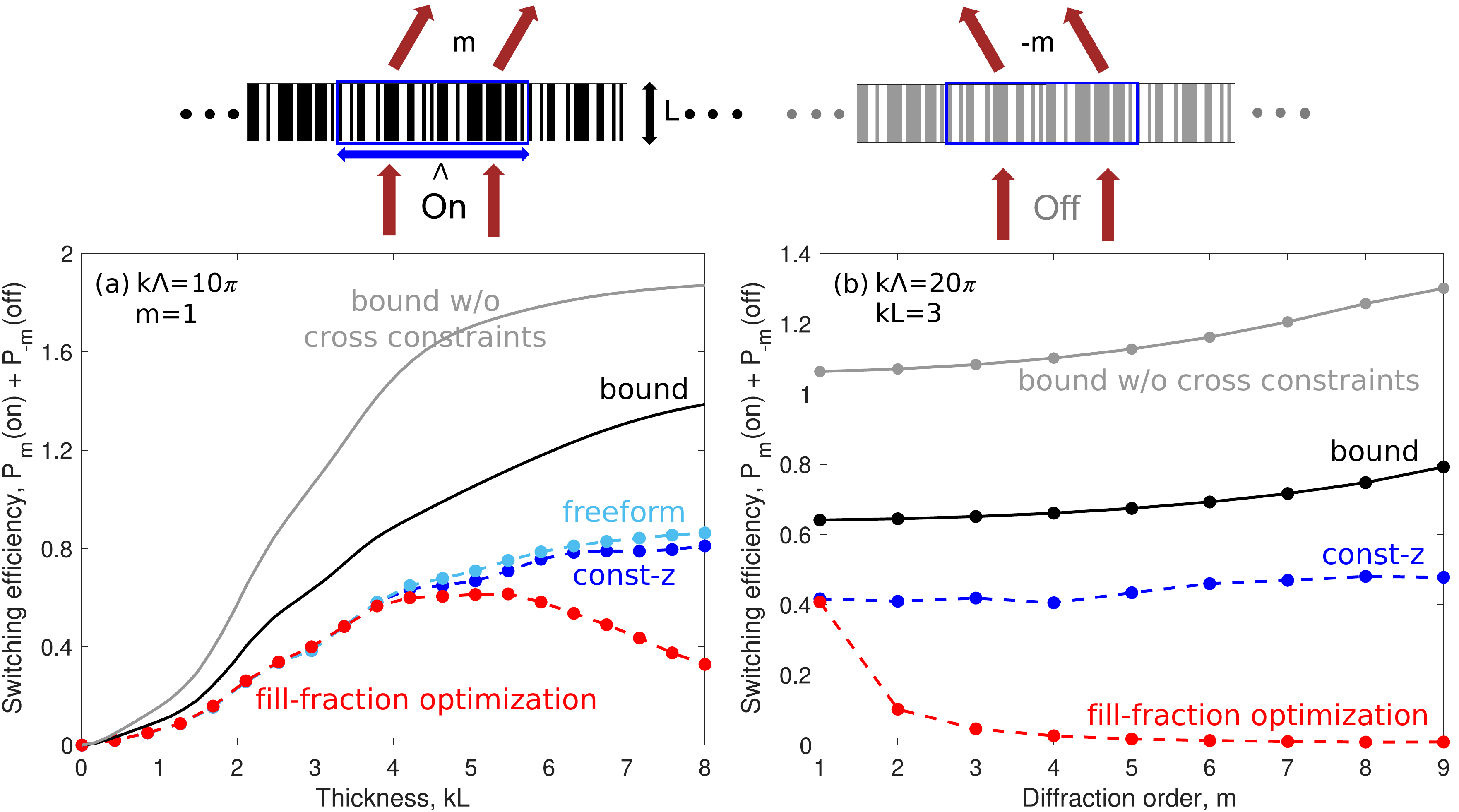} 
    \caption{(Top) Liquid-crystal-based beam beam-switching device. (a,b) Bounds on switching efficiency, both with and without cross-correlation constraints, compared to 3 different designs: simple grating structure with optimal fraction of material (fill-fraction optimization), lithography-friendly optimization with equal depth of air holes (const-z), and arbitrary permittivity at any point in the designable region (freeform). The switching efficiency is plotted against (a) thickness of the designable region for unit cell period $k\Lambda=10\pi$ and diffraction order $m=1$ and (b) diffraction order for unit cell period $k\Lambda=20\pi$ and and thickness $kL=3$. The switching efficiency is the sum of the power into diffraction orders $m$ and $-m$ as illustrated at the top, where the power diffracted into each state is normalized to 1 (so that the maximum switching efficiency is 2). The freeform designs are not shown in (b) because they are identical to const-z optimization in terms of switching efficiency at the (subwavelength) thickness of $kL=3$.}  
    \label{fig:2Dswitching} 
\end{figure*} 
		 
Another prototypical example of a multi-functional optical device is one in which a voltage is applied to change the refractive index and modulate the optical response. Here, we consider an example of liquid-crystal-based beam switching, in which the target functionality is to direct light in one direction for one voltage state, and another direction for another voltage state. The design of such unit cells could be useful not only for two-state switching in periodic structures, but also for many-state switching in a larger composition of varying unit cells~\cite{Patel1995,Lee2008}.

We consider periodic grating structures with the active designable medium comprising a liquid crystal (LC). (More realistic models with a LC filling a patterned material such as silicon can be similarly modeled~\cite{Chung2020b}.) For a period $\Lambda$ and grating thickness $L$ (as shown in the top of \figref{2Dswitching}), we consider any possible pattern of the LC material. We choose E7~\cite{Yang2010} as the LC material, which has refractive indices of about $1.7$ and $1.5$ in the voltage-on and off state respectively (for purposes of demonstration, we allow for a small $\Im n$ of 0.1 for both cases). Given a monochromatic field at normal incidence, we choose our objective to be the power switching efficiency, defined to be the sum of the power in the target directions for the voltage-on and voltage-off. We choose diffraction orders $m$ and $-m$, with angles $\theta_m = \sin^{-1}(2\pi m / k\Lambda)$ and $\theta_{-m} = -\theta_m$ with respect to the normal, to define the target directions for the on and off states, respectively. The power diffracted into the two states, $P_m$ and $P_{-m}$, are quadratic functions of the polarization field and naturally fit our bound framework. 

\Figref{2Dswitching}(a) shows bounds and designs for the liquid-crystal beam-switching problem with the unit cell taken to be about 5 free-space wavelengths wide ($k\Lambda = 10\pi+\delta$), and take $m=1$. (The constant $\delta = 0.1$ is added to avoid the singularities and infinite-$Q$ resonances possible via bound-state-in-continuum modes~\cite{Hsu2016}, which cause numerical instabilities.) Hence the angular deflection is small ($\pm 11.5^\circ$), and one might expect a perfect switching efficiency to be relatively easy to achieve. Yet the solid black line shows the bound computed using the cross-correlation constraints, which even for significant thickness ($kL = 8$) shows the possibility to reach only about $70\%$ of the maximum possible switching efficiency (which is 2). We also consider three possible design strategies: a fill-fraction optimization (red), in which the optimal fill fraction of a simple grating structure (fill fraction here specifies the fraction of the material occupying the unit cell) is computed, a ``const-z'' optimization (blue), in which the all air holes must have equal depth to be compatible with lithography, and a freeform optimization, in which the permittivity is allowed to take either material value at any point in the domain (teal). The freeform optimizations have the most degrees of freedom, and come closest to the bounds, while the fill-fraction optimizations are feasible at small thicknesses but deteriorate in quality at larger thicknesses. The freeform and const-z approaches both show similar trendlines to the computed bounds. By contrast, the bounds without cross-correlation constraints quickly approach 2, which is a trivial value that would suggest $100\%$ possible efficiency in both states. \Figref{2Dswitching}(b) isolates a single thickness $kL = 3$ for a larger unit-cell period $k\Lambda = 20\pi$ and sweeps over the target diffraction order $m$, with angular deflections increasing to $64^\circ$ for $m=9$. Perhaps surprisingly, the bound suggests that at this large unit-cell period, the switching efficiency can \emph{increase} as the angular of deflection increases. This is borne out by the const-z optimizations, which show a similar trend (though the noisiness of local optimizations makes the trend less clear than the bounds do).
	
\section{Extensions}

The essence of our approach is as follows: starting from a linear partial differential equation (PDE) that governs the response of field variables under multiple scenarios, one can transform the PDE to an integral equation in terms of polarization-field variables. Next, one can transform the linear but domain-dependent integral equations to quadratic, \emph{domain-independent} equations. Crucially, those quadratic constraints largely comprise cross-correlation equations that are necessary to describe multi-scenario problems. Upon reaching this QCQP, standard methods can be used to compute bounds via semidefinite programming. 

No part of our approach requires the linear PDE to be Maxwell's equations. And, indeed, this approach can be applied to applications such as quantum optimal control~\cite{Peirce1988,Werschnik2007,dAlessandro2007,Brif2010,Glaser2015,Zhang2021}, and potentially to applications in acoustics, elasticity, and more~\cite{Bendsoe2013}. Of course, within nanophotonics, the choice of an electric susceptibility and electric-field variable as written in \eqref{MaxDesProb} is not unique, and this approach can be applied for arbitrary materials and for electric and/or magnetic fields. 

Looking forward, a key opportunity is in the development of faster computational algorithms for solving the semidefinite programs that arise in this approach. Interior-point algorithms are highly effective but scale at least with the cube of the number of degrees of freedom of the design~\cite{ma2008some}. First-order methods such as ADMM~\cite{Boyd2010} have been shown to achieve near-linear scaling for certain instances of SDPs~\cite{Oliveira2018}, and would come with an additional benefit: integral operators have a special hierarchical rank structure ~\cite{Martinsson2019}, oft-utilized for fast multipole methods~\cite{greengard1987fast,coifman1993fast}, for example, that could provide significant further speed improvements of matrix multiplications with the integral operators.

In addition to improving SDP computation times, another avenue is to reduce the total number of constraints that are necessary. One approach, as developed in \citeasnoun{Kuang2020b}, is to iteratively choose only the constraints that are most violated by a given solution. Another approach, unique to the multi-scenario case, is to choose a minimal set of pairs of indices for which to impose the cross-correlation constraints: instead of enforcing cross-correlation across every pair, for example, enforcing it between pairs 1 and 2, then 2 and 3, and so on, until $N$ and $N-1$, would actually enforce exactly the same ``single-structure'' conditions but with $\sim N$ constraints instead of $\sim N^2$. The QCQP would be unchanged, though it is unclear whether the resulting SDP formulations would be. This is connected with a well-known opportunity in SDPs, in which at times it can be beneficial to impose ``redundant'' constraints in a quadratic formulation but lead to better SDP bounds~\cite{Anstreicher2000}. 

Connecting nanophotonics design problems to QCQPs in optimization theory offers a new lens with which to view them, and a new approach for understanding what is possible.
 
\bibliography{mf_bib}

\end{document}